\documentclass[doublecol]{epl2} % for 2 columns style without line numbers
\usepackage{amssymb,amsmath}

\title{DNA confined in a two-dimensional strip geometry}
\shorttitle{DNA confined in a 2D strip} %Insert here a short version of the title if it exceeds 70 characters

\author{Aiqun Huang \and Aniket Bhattacharya }
\shortauthor{A. Huang \etal}

\institute{                    
  Department of Physics - University of Central Florida, Orlando, Florida 32816-2385, USA\\
}

\abstract{
Semiflexible polymers characterized by the contour length $L$ and persistent length $\ell_p$ confined in a spatial region $D$
have been described as a series of ``{\em spherical blobs}'' and ``{\em deflecting lines}'' by de Gennes and Odjik  
for $\ell_p < D$ and $\ell_p \gg D$ respectively. Recently new intermediate regimes ({\em extended de Gennes} and  
{\em Gauss-de Gennes}) have been investigated by Tree {\em et al.} [Phys. Rev. Lett. {\bf 110}, 208103 (2013)]. 
In this letter we derive scaling relations to characterize these transitions in terms of universal scaled fluctuations
in $d$-dimension as a function of $L,\ell_p$, and $D$, and show
that the Gauss-de Gennes regime is absent and extended de Gennes regime is vanishingly small for polymers 
confined in a 2D strip. We validate our claim by extensive Brownian dynamics (BD) simulation which also reveals
that the prefactor $A$ used to describe the chain extension in the Odjik limit 
is independent of physical dimension $d$ and is the same as previously found by 
Yang {\em et al.}[Y. Yang, T.~W. Burkhardt, G. Gompper, Phys. Rev. E {\bf 76}, 011804 (2007)].
Our studies are relevant for optical maps of DNA stretched inside a nano-strip. }

\begin{document}

\maketitle

Conformations and dynamics of DNA inside a nanochannel have attracted considerable attention among various disciplines of 
science and engineering~\cite{Reisner_Review}. 
Important biomolecules, such as, chromosomal DNAs, or proteins whose functionalities are crucially dependent on the exact sequence of the 
nucleotides or amino acids  usually exist in highly compact conformations. 
By straightening these molecules on a two dimensional sheet~\cite{Schwartz_PNAS_2010,Krishnan,Dekker_PRL_2008} or inside a 
nanochannel~\cite{Reisner_PRL_2005,Schwartz_Labchip,Tegenfeldt_PNAS_2004,Craighead_PRL_2005,Reisner_PNAS_2010,Purohit_PLos_2011}
it is possible to obtain the structural details of these molecules.  
It is believed that a complete characterization of the DNA sequence for each individual and a proper understanding the role of 
genetic variations will lead to personalized medicine for diseases, such as, cancer~\cite{Mardis_Nature_2011}. DNA confined and stretched inside a nanochannel
offers significant promise towards this goal. Unlike, traditional sequencing using Sangers method~\cite{Sangers} which requires fragmentation and 
replication, analysis of a single DNA will be free from statistical errors and sequence gaps while reconstruction~\cite{Mardis_Nature_2011}. 
Naturally quests for 
efficient but low cost techniques have attracted considerable attention. Along with optical 
maps~\cite{Schwartz_PNAS_2010,Tegenfeldt_PNAS_2004}, recently DNA melting 
characteristics inside a nanochannel have been 
studied showing further promises~\cite{Reisner_PNAS_2010}. These recent experiments have generated renewed interests 
in theoretical and computational studies of confined polymers~\cite{Burkhardt_PRE_2007}-\cite{Ala-Nissila_MM_2013}.\par
Confined DNAs inside nanochannels often studied in high salt concentrations~\cite{Reisner_Review} where the charges of the individual nucleotides are heavily 
screened~\cite{Dorfman_MM_2011,Schwartz_Labchip}.
Besides, the resolution of optical studies set by the diffraction limit is typically of the order of 100 base pairs. Under these conditions 
a double-stranded DNA is often described as a worm-like chain 
(WLC)~\cite{Rubinstein} whose end-to-end distance 
$\langle R_{bulk}^2 \rangle = 2 \ell_p L \left(1-\frac{\ell_P}{L} \left[1- \exp \left(-L/\ell_P \right) \right]\right)$
interpolates from a rod ($\langle R_{bulk}^2 \rangle \sim L^2$ for $ L \ll \ell_p$) 
to a Gaussian coil ($\langle R_{bulk}^2 \rangle \sim 2L\ell_p$ for $L \gg \ell_p$).
However, for a very long chain eventually the excluded volume (EV) effect becomes important~\cite{Hsu_EPL_2010,Moon_1991}, 
and for $ L \gg \ell_p$ the end-to-end distance in $d$ dimensions should be characterized by the bulk conformation of a 
swollen semiflexible chain~\cite{Pincus_MM_1980,Nakanishi_1987}
\begin{equation}
\sqrt{\langle R_{bulk}^2 \rangle} = a\left(\frac{L}{a}\right)^{\frac{3}{d+2}}\left(\frac{\ell_p}{a}\right)^{\frac{1}{d+2}},
\label{r1n}
\end{equation}
where $a$ is the effective width of the chain.
It is noteworthy that while in 3D there is a broad Gaussian regime for $L \gtrapprox \ell_p$~\cite{Hsu_EPL_2010,Moon_1991} before EV effects become important, 
in two dimensions (2D) the intermediate Gaussian regime is absent
due to severe dominance of the EV effect~\cite{Hsu_EPL_2011,Huang_EPL_2013}. \par
Recently confined polymers in rectangular, cylindrical and triangular channels have been studied by 
several groups~\cite{Burkhardt_PRE_2007}-\cite{Ala-Nissila_MM_2013}.
One first sees the effect of the confinement (described by the length of the cross section of the channel $D$) for $D < R_g$, where $R_g$ 
is the radius of gyration of the chain. This limit has been identified as the Flory-de Gennes regime where chain conformations can be 
described as a series of spherical blobs of size $D$~\cite{Daoud_1977,deGennes}. Further decrease of 
the ratio $D/R_g$ first leads to an extended de Gennes regime with anisotropic blobs followed by 
a Gaussian regime analogous to the 3D bulk case, which has been referred as the Gauss-de Gennes regime~\cite{Dorfman_PRL_2013a}; for extreme confinement 
when $\ell_p \gg D$, the blob picture 
breaks down and  the chain enters into the Odjik regime~\cite{Odijk_JPSC_1977,Odijk_MM_1983} where the chain conformations are described as a series of straight segments deflected 
from the confining wall~\cite{Odijk_PRE_2008}. While both de Gennes and Odjik regimes are well established, characteristics of the transition regions (the extended de Gennes and the Gaussian)
have been the main goal of several recent studies~\cite{Odijk_PRE_2008,Dorfman_MM_2011,Dorfman_PRL_2013a,Doyle_SM_2012}. However, the extension of the confined 
DNA in the extended de Gennes regime has been determined to be the same as in the de Gennes regime by minimizing the 
free energy~\cite{Reisner_Review,Dorfman_MM_2011,Doyle_SM_2012}, so making a difference of these two regimes has been either difficult or not 
evident~\cite{Dorfman_MM_2011,Dorfman_PRL_2013a}. \par
Confined chains inside 3D nano-channels exhibit analogous regimes as found in their respective bulk counterparts~\cite{Dorfman_PRL_2013a}.  
In this letter we study confined DNA in a 2D strip geometry. As mentioned before, that unlike in 3D, the bulk Gaussian regime  
does not exist for semiflexible chains in 2D~\cite{Hsu_EPL_2011,Huang_EPL_2013}. Therefore, one wonders, if regimes of confined DNA in 2D will follow their corresponding 
bulk counterpart. The second motivation comes from the observation that 
the Flory exponent in 2D (0.75) is significantly larger than the corresponding exponent in 3D (0.588) which implies that a chain is more elongated in a 2D strip rather than in a tube of the same width $D$. Therefore, the elongation would be more profitable by further reducing the physical dimension of the region. 
Finally, in the Odjik limit, prior theoretical and numerical results~\cite{Burkhardt_PRE_2007} have indicated that the prefactor $A$ in the expression for the 
chain elongation (see Eqn.~\ref{Odijk_X}) is 
nearly independent of the shape of the nanochannel. By studying elongation along a 2D strip, we further observe that this constant 
is almost the same as the values in 3D indicating that this constant is independent of the spatial dimension. 
While in 3D the extended de Gennes limit is somewhat controversial, 
we provide scaling arguments for a 2D strip and validate by carrying out BD simulation that the extended de Gennes regime is vanishingly small. 
This result along with the absence of a Gaussian regime in a 2D strip geometry implies that a 2D strip is a cleaner system 
to study a stretched chain as the conformations interpolate between de Gennes and Odjik regimes only, and therefore,
is another reason to think about designing DNA elongation experiments inside a 2D strip.\par   
$\bullet$~{\em de Gennes Regime:}~The starting point of our theoretical analysis is the ansatz 
for the normalized free energy ${\mathcal F}/k_BT$ of 
confinement along a tube axis first proposed by Jun, Thirumalai, and Ha~\cite{Thirumalai_PRL_2008}, 
later used for a square channel~\cite{Dorfman_MM_2011} and a slit~\cite{Doyle_SM_2012}, and is given by
\begin{equation}
{\mathcal F}/k_BT = \, \frac{X^2}{(L/L_{blob})\,D^2} \, + \, D\frac{(L/L_{blob})^2}{X}, \label{jun}\\
\end{equation}
and the expression for the end-to-end distance of a swollen semiflexible chain as given by Eqn.~\ref{r1n}. 
Here $X$ is the extension along the tube/strip axis, and $L_{blob}$ the contour length of the chain 
in a blob~\cite{Daoud_1977,deGennes}, $k_B$ is the Boltzmann constant, and $T$ is 
the temperature. The dimension dependence comes from the chain statistics for $L_{blob}$ (Eqn.~\ref{r1n}). In order to contrast the results for polymers confined
in a 2D strip with those 
for cylindrical, square, and rectangular channels, in the following we derive expressions in terms of $d$ spatial dimensions ($d=2$ for a strip and $d=3$ for a tube). 
By differentiating Eqn.~\ref{jun} with respect to $X$, one can easily check (i) $X = Dn_{blob}$, where $n_{blob} = L/L_{blob}$ is the number of blobs, (ii) 
${\mathcal F}/k_BT \sim n_{blob}$, and (iii)  ${\mathcal F}/k_BT \sim L$. For the de Gennes regime monomers inside the blob are described by 
the conformation of a swollen chain either in $d=2$ (strip) or $d=$3 (tube), so that $D = {L_{blob}}^{3/d+2}{\ell_p}^{1/d+2}a^{d-2/d+2}$ (Eqn.~\ref{r1n}). It is then 
easy to check that the elongation is given by
\begin{equation}
\langle X \rangle_\mathrm{de \, Gennes} = Dn_{blob} = L \left(\frac{D}{a}\right)^{\frac{1-d}{3}}\left(\frac{\ell_p}{a}\right)^{\frac{1}{3}}. 
\label{deGx}
\end{equation}
Likewise, the second derivative of Eqn.~\ref{jun} gives the effective stiffness constant $k_{\mathrm{eff}}$ 
for the DNA polymer~\cite{Reisner_Review,Reisner_PRL_2005} 
under confinement, so that the longitudinal fluctuation of the extension $\langle\sigma^2\rangle$ can be obtained as
\begin{equation}
\langle\sigma^2\rangle=\frac{k_BT}{k_{\mathrm{eff}}} = La\left(\frac{\ell_p}{a}\right)^{\frac{1}{3}}\left(\frac{D}{a}\right)^{\frac{4-d}{3}}.  
\label{sigma}
\end{equation}
$\bullet$~{\em Extended de Gennes Regime:}~It was argued~\cite{Reisner_Review,Odijk_PRE_2008,Dorfman_MM_2011} that the scaling relation 
Eqn.~\ref{r1n} for each spherical blob in de Gennes regime only holds true when the channel size $D$ and chain length $L$ both are above 
certain critical values $D_{**}$ and $L_{**}$ respectively to be determined in the following manner. When $D < D_{**}$ the EV repulsion becomes less significant resulting in a local ideal 
chain behavior in each blob, while strong enough to sustain the global picture of linearly ordered blobs, each turning into an 
ellipsoid characterized by its major axis $H$ (and of volume $\sim D^{d-1}H$) 
along the long axis of the nanochannel~\cite{comment1}. The critical 
length $L_{**}$ and the critical channel width $D_{**}$ can be obtained by 
equating the size of an ideal chain and a Flory coil in the bulk:  
$D_{**} \simeq \, \left( L_{**}\ell_p \right)^{1/2} \simeq \, \ell_p\, ^{\frac{1}{d+2}} \, L_{**}\, ^{\frac{3}{d+2}}$, 
%\begin{equation}
%D_{**} \simeq \, \left( L_{**}\ell_p \right)^{1/2} \simeq \, \ell_p\, ^{\frac{1}{d+2}} \, L_{**}\, ^{\frac{3}{d+2}}, \label{trans1},
%\end{equation}
%From Eqn.~\ref{trans1} we get
from which we get
\begin{equation}
L_{**} \simeq \, a\left( \frac{\ell_p}{a} \right)^{\frac{d}{4-d}} {\rm ~~~~~and~~~}
D_{**} \simeq \, a\left( \frac{\ell_p}{a} \right)^{\frac{2}{4-d}}. 
\label{L**}
\end{equation}
Notice that $L_{**} \simeq l_p^3 a^{-2}$, $D_{**} \simeq  l_p^2 a^{-1}$ in 3D while $L_{**} \simeq D_{**} \simeq l_p$ in 2D. 
Both ideal and EV effects coexist in this regime~\cite{Odijk_PRE_2008}. This balance of ideal and EV behavior is obtained by setting  \
$a^{d-2}L_{blob}^2/HD^{d-1}=1$ from which the length $H$ can be obtained as follows:
%$H = \sqrt{L_{blob}\ell_p} = a^{d-2}\frac{L_{blob}^2}{D^{d-1}}$%
\begin{equation}
H = \sqrt{L_{blob}\ell_p} = a^{d-2}\frac{L_{blob}^2}{D^{d-1}} \label{exdeg1}
\end{equation}
Denoting $L_{blob}$ as $L_{ellip}$, from Eqn.~\ref{exdeg1} we get
\begin{equation}
L_{ellip}  = {\ell_p}^{\frac{1}{3}}\left(\frac{D^{d-1}}{a^{d-2}}\right)^{\frac{2}{3}} {\rm ~and~} 
H = \ell_p^ {\frac{2}{3}}\left(\frac{D^{d-1}}{a^{d-2}}\right)^ {\frac{1}{3}} \label{L*}. 
\end{equation}
Replacing $ D \rightarrow H$, $ L_{blob} \rightarrow L_{ellip}$ in Eqn.~\ref{jun} and minimizing with respect to $X$, we get $X = H(L/L_{ellip})$, 
and substituting $L_{ellip}$ and $H$ by Eqn.~\ref{L*} we obtain 
Eqn.~\ref{deGx}. This completes the proof that  
the de Gennes regime and the extended de Gennes regime can not be differentiated from the elongation of the chain. \par 
However, by repeating the same procedure we note, unlike Eqn.~\ref{sigma} in the extended de Gennes regime the 
fluctuation is different and is given by $\langle\sigma^2\rangle=L\ell_p$. 
Therefore, the de Gennes regime and the extended de Gennes regime can be differentiated by measuring the characteristic fluctuations in their respective 
chain extensions~\cite{Dorfman_MM_2011}. The lower bound $D_*$ of the extended de Gennes regime where it merges with the Gauss-de Gennes regime, 
following Odijk's scaling analysis ~\cite{Odijk_PRE_2008} (which is also valid in 2D) is given by $D_{*} \simeq c\ell_p$,  where the prefactor $c \gtrapprox 1$ (in \cite{Dorfman_MM_2011} 
it was found to be $\approx 2$). Using Eqn.~\ref{L**} we note that 
while the range for the extended de Gennes regime being $[D_*, D_{**}]=[c\ell_p,\, \ell_p^2]$ is broad in 3D, 
it would be either very narrow or vanishingly small to be observed in $[c\ell_p, \,D_{**} \simeq \ell_p]$ in 2D. \par
%%%%%%%%%%%%%%%%%%%%%%%%%%%%%%%%%%%%%%%%%%%%%%%%%%%%%%%%%%%Gauss begins%%%%%%%%%%%%%%%%%%%%%%%%%%%%%%%%%%%%%%%%%%%%%%%%%%%%%%%%%%%%%%%%%%%%%%%%%%
%%%%%%%%%%%%%%%%%%%%%%%%%%%%%%%%%%%%%%%%%%%%%%%%%%%%%%%%%%%%%%%%%%Gauss ends%%%%%%%%%%%%%%%%%%%%%%%%%%%%%%%%%%%%%%%%%%%%%%%%%%%%%%%%%%%%%%%%%%%%
$\bullet$~{\em Gauss-de Gennes Regime:}~Upon further decrease of the confining region, for $D < D_*$ the EV effect plays no role,  
and the DNA behaves as a Gaussian chain~\cite{Odijk_PRE_2008,Dorfman_PRL_2013a}, so that 
$D=\left(L_{blob}\ell_p\right)^{1/2}$ ~\cite{Dorfman_PRL_2013a}. Then according to Eqn.~\ref{jun}, we have the extension
\begin{equation}
\langle X \rangle_\mathrm{Gauss-de Gennes}=L\frac{\ell_p}{D},  \label{Gauss-de Gennes}
\end{equation}
which holds both in 2D and 3D. It is easy to check that in this regime the fluctuation $\langle\sigma^2\rangle=L\ell_p$, the 
same as in the extended de Gennes regime. \par
While Eqn.~\ref{Gauss-de Gennes} has been recently tested to be true for 3D~\cite{Dorfman_PRL_2013a} channels, 
similar studies have not been done for confined polymers in 2D strips. Considering the absence of Gaussian regime for a bulk 2D swollen 
chain~\cite{Hsu_EPL_2011,Huang_EPL_2013} one wonders if this new Gauss-de Gennes phase will 
be observed in a 2D strip. The universal fluctuations from our BD simulation studies (Fig.~\ref{sigmaeps}) will provide conclusive evidence 
for the absence of a Gaussian regime inside a 2D strip. \par
$\bullet$~{\em Odjik Regime:}~
For $\ell_p \gg D$  Odijk \cite{Odijk_JPSC_1977,Odijk_MM_1983} argued that the chain deflects back and forth off the wall with a deflection 
length of $\lambda \simeq (\ell_pD^2)^{1/3}$, and the extension of the confined polymer can be written as~\cite{comment2}
\begin{equation}
\langle X \rangle_\mathrm{Odijk}=L\left[ 1 - A \left( \frac{\ell_p}{D} \right)^{-\frac{2}{3}} \right],  \label{Odijk_X}
\end{equation}
where $A$ is a ``universal''\cite{universal} prefactor~\cite{Burkhardt_PRE_2007,Burkhardt_PRE_2010}. In this limit it is easy 
to check that the the free energy and fluctuations in chain length both in 2D and 3D are given by 
\begin{equation}
F/k_BT= B \frac{L}{(\ell_pD^2)^{1/3}} {\rm ~~and~~} \langle\sigma^2\rangle=\frac{LD^2}{\ell_p}  \label{F_Odijk}.
\end{equation}
$\bullet$~{\em Brownian dynamics(BD) simulation results:}~To provide further support to our scaling analyses 
we have performed Brownian dynamics (BD) simulation with a bead-spring model for a swollen chain having 
pairwise repulsive Lennard-Jones (LJ) interaction between any two monomers (excluded volume), a finitely extensible nonlinear elastic (FENE) potential 
between the successive beads (elastic bond energy), and a 
\begin{figure}[ht!]                
\centering
\includegraphics[width=0.9\columnwidth]{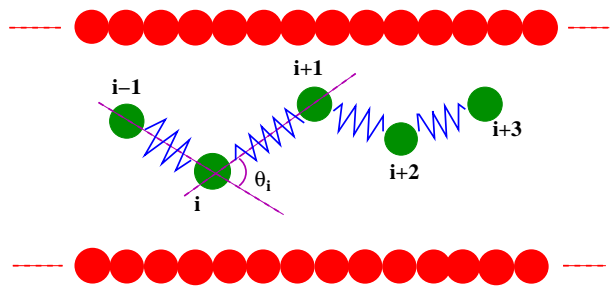}
\caption{\small Bead-spring model of a 2D polymer confined in a 2D channel.}
\label{model}
\end{figure}
three-body potential $U_\mathrm{bend}=\kappa (1 - \mathrm{cos}\theta_i)$, where $\theta_i$ (Fig.~\ref{model}) is the angle between two consecutive 
bonds, and the parameter $\kappa=\frac{1}{2}k_BT\ell_p$ is a measure of the chain stiffness proportional to the chain persistence length $\ell_p$. The DNA-wall interaction is also modeled as LJ. We observed that during simulation the average bond length stays at 0.97 with a 
fluctuation less than 0.2\%, and the bending potential hardly affects the bond length. 
By monitoring $\langle \mathrm{cos} \theta \rangle$ we also find that 
$\ell_p=-1/\mathrm{ln}\left( \langle \mathrm{cos} \theta \rangle\right) \equiv 2\kappa/k_BT$ to be the   
same as in a WLC~\cite{Huang_EPL_2013}. Numerical 
integration of the equation of motion with respect to time in the canonical ensemble was done according to the algorithm developed 
by Gunsteren and Berendsen \cite{Gunsteren}. 
In our simulation we have used reduced units of length, time, and temperature to be $a$, $a\sqrt{\frac{m}{\epsilon}}$, $\epsilon/k_B$, respectively. We have chosen a large number of combinations of $256 \le N \le 1024$, 
the chain persistence length $2 \le \ell_p \le 270 $ (by varying $\kappa$ from 1 to 160 ), 
and the strip width $D$ = 18, 36, and 80 such that the ratio $l_p/D$ is in the window  $0.025 \le l_p/D \le 15 $ and $ 1 \le L/\ell_p \le 400$.  With these choices we cover experimental study scales (the commonly used $\lambda$ DNA in experiments has a contour length $L=16.5 \mu$m with a persistence length $\ell_p \simeq 50$ nm, and the channel diameter ranges between 10 nm -200 nm~\cite{Reisner_Review}) and fully interpolate from the de Gennes limit to the Odijk limit. The confined chains were 
equilibrated for several Rouse relaxation time before data were collected over a span of 10 - 25 Rouse relaxation time to ensure 
convergence. \par
\begin{figure}[ht!]                
\centering
\includegraphics[width=0.92\columnwidth]{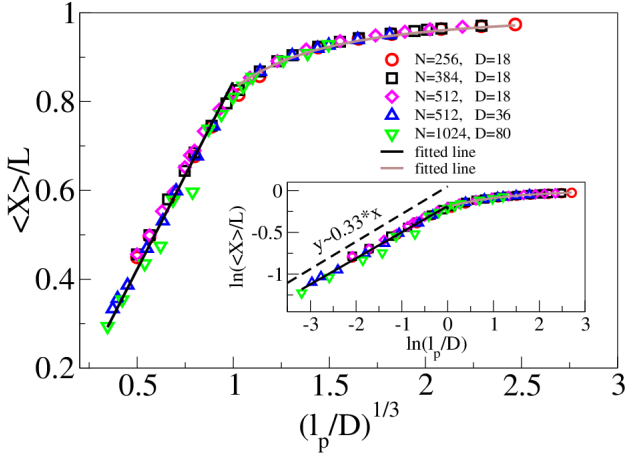}
\caption{\small Dimensionless chain extension $\langle X \rangle / L$ as a function of $(\ell_p/D)^{1/3}$ for various combination of chain length $N$, 
persistence length $\ell_p$, and width $D$ of the confining strip. The inset is the log-log plot $\langle X \rangle / L$ as a function of $\ell_p/D$ showing excellent data collapse with initial slope of 1/3 for $\ell_p \leq D$ verifying Eqn.~\ref{deGx}.}
\label{RPD}
\end{figure}

\begin{figure}[ht!]                
\centering
\includegraphics[width=0.92\columnwidth]{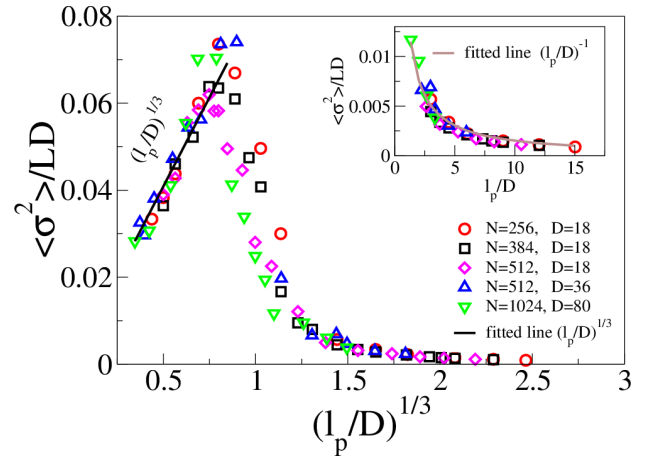}
\caption{\small Normalized fluctuation $\langle \sigma^2 \rangle /LD$ as a function of $(\ell_p/D)^{1/3}$ for the same combinations of $N$, 
$\ell_p$, and $D$ as in Fig.~\ref{RPD}. The inset shows $(\ell_p/D)^{-1}$ dependence in the Odijk limit. } 
\label{sigmaeps}
\end{figure}

 
Fig.~\ref{RPD} shows the normalized chain extension. All the data for many combinations of $L,\ell_p$, and $D$ collapse onto one master curve and shows a 
smooth transition from de Gennes regime to Odijk regime, which also indicates the absence of Gauss-de Gennes regime predicted by Eqn.~\ref{Gauss-de Gennes}.
For $\ell_p \leq D$ excellent linear fit of $\langle X \rangle /L \sim  (\ell_p/D)^{1/3}$ validates theoretical prediction 
of de Gennes regime (Eqn.~\ref{deGx}). For  $\ell_p > D$ 
we used Eqn.~\ref{Odijk_X} to fit the data and  
the prefactor is determined to be $A_{strip}=0.171$. With prior reported values for this prefactor $A_{square}=0.183$ and $A_{cylin}=0.170$ 
~\cite{Burkhardt_PRE_2007,Burkhardt_PRE_2010}) it indicates that the constant $A$ has little dependence on the physical spatial dimension and nearly universal, being consistent
with the fact that one can show the validity of Eqn.~\ref{Odijk_X} in both 3D and 2D. In the log-log plot shown in the inset of Fig.~\ref{RPD}, the $1/3$ power law dependence in the de Gennes regime expands to $l_p/D \simeq 1$ and the scaling relation Eqn.~\ref{Gauss-de Gennes} in Gauss-de Gennes regime is not seen at all.
Furthermore, around $\ell_p/D \simeq 1$ we find that both Eqn.~\ref{deGx} as well as Eqn.~\ref{Odijk_X} give {\em almost the same value for the extension, which shows that a description by Eqn.~\ref{Gauss-de Gennes} is not necessary indicating that there is no Gauss-de Gennes regime between them}.\par
\begin{figure}[ht!]                
\centering
\includegraphics[width=0.92\columnwidth]{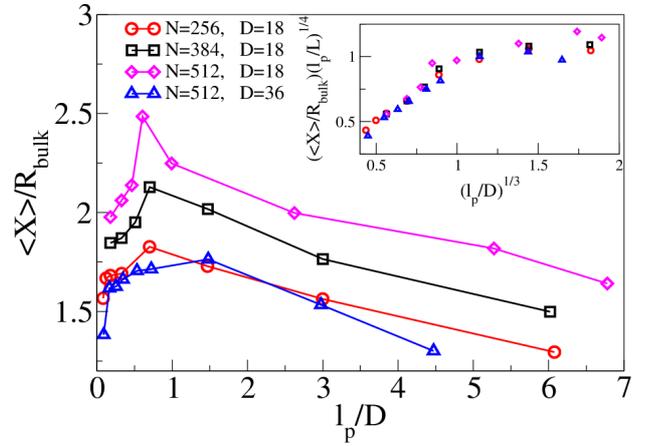}
\caption{\small Plot of the normalized extension by the end-to-end distance $R_\mathrm{bulk}$ in the bulk. The inset shows plot of Eqn.~\ref{peak_left}.} 
\label{cees}
\end{figure}

%%++++++++++++++++++++++++++++++++++++++++++++++++++++++++++++++++++++++++++++++++++++++++++++++
We also observe another interesting feature by plotting  chain extensions  $\langle X \rangle$ 
normalized by the corresponding bulk end-to-end distance $R_\mathrm{bulk}$ as a function of $\ell_p/D$ which exhibits a peak for each curve as shown in Fig.~\ref{cees}.
This peak can be reconciled by noting that the normalized extensions in the de Gennes and Odijk limits can be expressed as
%%%%%%%%%%%%%%%%%%%%%%%%%%%%%%%%%%%%%%%%%%%Fig.3%%%%%%%%%%%%%%%%%%%%%%%%%%%%%%%%%%%%%%%%%%%%%%%%%%%%
\begin{subequations}
\begin{align}
\frac{\langle X \rangle_\mathrm{de \, Gennes} }{R_\mathrm{bulk}}\, &= \, \left( \frac{L}{\ell_p} \right)^{1/4} \left( \frac{\ell_p}{D} \right)^{1/3} = \tilde{L}\tilde{\ell_p}^{1/12}, \label{peak_left} \\
\frac{\langle X \rangle_\mathrm{Odijk} }{R_\mathrm{bulk}}\, &= \,\left( 1-A{\tilde{\ell_p}}^{-2/3} \right)\tilde{L}^{1/4}\tilde{l_p}^{-1/4},
\label{peak_right}
\end{align}
\end{subequations}
%%%%%%%%%%%%%%%%%%%%%%%%%%%%%%%%%%%%%%%%%%%%%%%%%%%%%%%%%%%%%%%%%%%%%%%%%%%%%%%%%%%%%%%%%%%%%%%%
where we have used $D$ as the unit of length (data points in each curve in Fig.~\ref{cees} have the same $D$) so that $\tilde{L} = L/D$ and $\tilde{\ell_p}=\ell_p/D$ 
respectively. Eqns.~\ref{peak_left} and \ref{peak_right} readily follow from Eqns.~\ref{r1n}, \ref{deGx}, and ~\ref{Odijk_X} respectively. One notices as $\ell_p/D$ increases, 
{\em i.e.} $\tilde{\ell_p}$ increases, and the extreme left and right side of the peak 
correspond to de Gennes and Odijk limits respectively. But from Eqn.~\ref{peak_left} and \ref{peak_right} we note that for small values of $\tilde{\ell_p}$ the normalized extension 
increases as $\sim \tilde{l_p}^{1/12}$ (de Gennes limit) whereas, for large values of $\tilde{\ell_p}$ the normalized extension decreases as $\sim \tilde{l_p}^{-1/4}$ (Odijk limit), which implies that for finite 
extension of a chain, the normalized extension will exhibit a maximum as a function of $\ell_p/D$. It is also noteworthy that this maximum occurs for $\ell_p/D \sim 1$ at the confluence
of de Gennes and Odijk limit. 
This description is also consistent with the critical channel 
width $D_* \simeq 2\ell_p$ which marks the onset of Odijk regime. We also plotted Eqn.~\ref{peak_left} 
which is only valid in de Gennes regime (i.e., for $\ell_p < D$) at the inset of Fig.~\ref{cees} showing data collapse similar to Fig.~\ref{RPD}.\par
We now show simulation results for the fluctuation in the chain extensions and compare these results with the theoretical predictions. 
According to Eqn.~\ref{sigma} and Eqn.~\ref{F_Odijk} the normalized fluctuation $\langle \sigma^2 \rangle /LD$ scales as $(\ell_p/D)^{1/3}$ and $(\ell_p/D)^{-1}$  
in the de Gennes and Odjik limits respectively.  Indeed we find in Fig.~\ref{sigmaeps} that the 
fluctuation grows as $\left(\ell_p/D\right)^{1/3}$ in the de Gennes regime until $\ell_p \simeq 0.5D$ 
when it enters the Odijk limit and decays as $\left(\ell_p/D\right)^{-1}$ (inset). 
It is reassuring to note that since both the extended de Gennes and the Gauss-de Gennes regimes do not occur inside a 2D strip, in Fig.~\ref{sigmaeps} 
we do not see any intermediate regime where $\langle \sigma^2 \rangle /LD \sim \ell_p/D$, the characteristic fluctuations of both the extended de Gennes as well as the Gauss-de Gennes regimes.
The excellent data collapse for same combinations of $N$, $\ell_p$, and $D$ as in Fig.~\ref{RPD} 
and the sharp peak signifies the onset of a transition from the de Gennes regime to the Odijk 
regime.\par
To summarize, in this letter we have provided a generalized scaling theory 
of confined DNA in $d$-dimensions and compared/contrasted the behavior 
in 2D with those in 3D reported recently in the literature. We validate the scaling analyses by   
BD simulation where we identify each regime from 
excellent data collapse for the characteristic universal dimensionless extensions and fluctuations in terms of the dimensionless parameter $\ell_p/D$.
From the scaling analysis and results from BD simulation reported in this letter, and prior work for 3D cylindrical and square channels, 
we concur that the different regimes of confined polymers 
follow their corresponding regimes in the bulk. 
We find that for a 2D strip, the Gaussian regime is absent and the extended de Gennes regime is vanishingly small, so that 
the chain conformations inside the channel are described either by the de Gennes or by the Odjik regime.  
Thus the chain conformations for a straightened DNA inside a 2D strip are cleaner than for those of the 3D cylindrical and square geometries. 
Therefore, we believe that this work will motivate further experimental and theoretical work to study confined DNA inside nano-strips.

\acknowledgments
The research has been partially supported by the UCF Office of Research \& Commercialization and 
the UCF College of Sciences SEED grant. We thank for the reviewers for constructive comments on the manuscript.
%Insert here the text.


\begin{thebibliography}{99}
\bibitem{comment2}This result is valid both in two and three dimensions as can be seen fron the expansion 
of various averages for worm like chains near the rod limit~\cite{Fuji,Yamakawa}.
\end{thebibliography}
\end{document}